\documentclass[sigconf]{acmart}

\copyrightyear{2020} \acmYear{2020} 
\setcopyright{rightsretained} \acmConference[SMSociety '20]{International Conference on Social Media and Society}{July 22--24, 2020}{Toronto, ON, Canada}\acmBooktitle{International Conference on Social Media and Society(SMSociety '20), July 22--24, 2020, Toronto, ON, Canada}\acmDOI{10.1145/3400806.3400813}\acmISBN{978-1-4503-7688-4/20/07}

\usepackage{natbib}
\usepackage{url}
\usepackage{balance}
\begin{document}

\title[Characterizing Online Vandalism]{Characterizing Online Vandalism: A Rational Choice Perspective}
\author{Kaylea Champion}
\affiliation{%
\institution{University of Washington}
\city{Seattle}
\state{WA}
\postcode{98195}
\country{USA}
}
\email{kaylea@uw.edu}

\renewcommand{\shortauthors}{Champion}

\begin{abstract}

What factors influence the decision to vandalize? 
Although the harm is clear, the benefit to the vandal is less clear. 
In many cases, the thing being damaged may itself be something the vandal uses or enjoys. 
Vandalism holds communicative value: perhaps to the vandal themselves, to some audience at whom the vandalism is aimed, and to the general public. Viewing vandals as rational community participants despite their antinormative behavior offers the possibility of engaging with or countering their choices in novel ways. Rational choice theory (RCT) as applied in value expectancy theory (VET) offers a strategy for characterizing behaviors in a framework of rational choices, and begins with the supposition that subject to some weighting of personal preferences and constraints, individuals maximize their own utility by committing acts of vandalism. This study applies the framework of RCT and VET to gain insight into vandals' preferences and constraints. Using  a mixed-methods analysis of Wikipedia, I combine social computing and criminological perspectives on vandalism to propose an ontology of vandalism for online content communities. I use this ontology to categorize 141 instances of vandalism and find that the character of vandalistic acts varies by vandals' relative identifiability, policy history with Wikipedia, and the effort required to vandalize. 
\end{abstract}



\maketitle    
                        
\section{Introduction}
Vandalism offers asymmetric power to the vandal. The object being damaged may originate in the effort of a powerful entity such as a large institution or a government or a large group of well-meaning people. A vandal's momentary action with a hammer or pen can be profoundly disruptive. 
In physical environments, communities may expend substantial resources to restore vandalized property: public works must be rebuilt, damaged artwork may be irretrievably lost, and time and money spent removing defacement could be deployed in other ways.

Online communities also struggle with vandalism. Automation in digitally-mediated environments can aid vandals who use tools to vandalize with incredible speed and volume \cite{geigerWhenLeveeBreaks2013}.  Although automated tools support rapid restoration of the digital work to its original state---often within minutes \cite{alkharashiVandalismCollaborativeWeb2018}, automated vandal fighting is not without costs. It requires development, invocation, and maintenance of tools and it leads to impersonal messages and improper rejection of contributions which has been implicated in a decline of well-intentioned newcomers \cite{halfakerRiseDeclineOpen2013}, a trend which requires substantial effort to remediate \cite{halfakerDonBiteNewbies2011, morganEvaluatingImpactWikipedia2018}.
The digital nature of online vandalism does not mean it is victimless---besides offending and burdening hard-working creators, maintainers, and moderators \cite{geigerWorkSustainingOrder2010}, the general public suffers as well. 
On highly visited websites, some information-seekers are likely to read the damaged page regardless of how quickly vandalism is removed. 

Given the social and individual cost and the difficult to discern source of benefit to the vandal, we may ask why it occurs at all. Competing theories in criminology seek to explain the motivations for and causes of crime, ascribing criminal behavior to such factors as lack of impulse control, lack of morals, or to societal failure. Alternatively, rational choice theory proposes that behaviors are the product of rational choices. In order to apply rational choice theory to vandalism, this project seeks to understand vandal decision-making in terms of preferences and constraints. Building on past examinations of Wikipedia contributors, we examine vandalism from four groups: users of a privacy tool, those contributing without an account, those contributing with an account for the first time, and those contributing with an account but having some prior edit history \cite{tranTorUsersContributing2020, championForensicQualitativeAnalysis2019}. 

This paper makes three contributions: I describe the extension of rational choice theory and value expectancy theory to Wikipedia, I offer an extension to the existing ontologies for characterizing online vandalism, and I demonstrate the application of both rational choice theory and this vandalism ontology in a preliminary series of findings. The paper is structured as follows. I offer background on the topic of vandalism and theories of rational choice in §\ref{sec:background}. I describe the comparison groups in §\ref{sec:data} and my approach to analysis in §\ref{sec:methods}. I share the results of classifying vandalism and examining the relative prevalence of each type in §\ref{sec:results} with limitations as described in §\ref{sec:limitations}. In §\ref{sec:discussion} I explore the implications of these findings, as well as how these results can inform design and policy in online communities. I conclude in §\ref{sec:conclusion}.

\section{Background}
\label{sec:background}

Wikipedia calls itself ``the free encyclopedia that anyone can edit'' and seeks to present encyclopedic information from a neutral point of view, with freely usable, editable, and distributable content, developed by a respectful community with no firm rules.\footnote{\url{https://en.wikipedia.org/wiki/Wikipedia:Five_pillars}} Like a public park, Wikipedia is a public good---defined as non-rivalrous in that one person's use does not consume it and non-excludable in that it is open to everyone. Technological advances have allowed for the collaborative development of new public goods through a process called commons-based peer production in which self-directed volunteers contribute according to their interests, whether small and casual or vast and serious \cite{benklerWealthNetworksHow2006}. However, in opening editing power to almost anyone, Wikipedia opens itself up to the possibility that not all participants engage in good faith to develop a high-quality, neutral encyclopedia. Instead, some contributors seek to do harm. 

As one of the top 10 most visited sites on the internet, damage to Wikipedia holds the potential to harm information seekers in multiple ways. Vandalism that deletes or disrupts a page will block access to information. Insertion of deliberate misinformation can generate rumors or distort understanding of a topic. Vandalism can expose the general public to threats or harassment. 

Other vandalism in Wikipedia is targeted to damage the production community itself, such as threatening and harassing contributors and administrators, wasting volunteer time by spamming requests for help when no help is needed, or by disrupting the behaviors of automated tools.

Some Wikipedia vandals write their names, add jokes, or make rude comments. These kinds of contributions are also characterized as vandalism in Wikipedia,\footnote{\url{https://en.wikipedia.org/wiki/Wikipedia:Vandalism}} although the content of the text is in some cases indistinguishable from what in the physical world would be called graffiti. Although graffiti has long been analyzed as a source of meaning in contexts ranging from ancient Pompeii to modern university library bathrooms \cite{dombrowskiWallsThatTalk2011}, the addition of unwelcome or disruptive contributions to Wikipedia is contrary to its use, making vandalism an apt description regardless of text. The Wikipedia community has a tradition of joking about persistent vandalism.\footnote{For example, this humorous essay describes categories of vandals as `typing students', `the curious', `critics', `men with big penises', `cheerleaders', and `friends of gays', the last being a group of people who the essay calls ``overly proud friends'' of gay and lesbian people, i.e. those whose vandalism consists of texts like ``[name] is gay.'' The essay is available at: \url{https://meta.wikimedia.org/wiki/Friends_of_gays_should_not_be_allowed_to_edit_articles}} 

Just as volunteers built the public good that is Wikipedia, they work to defend its utility from damage. However, countering vandalism in Wikipedia is a substantial undertaking. Vandal-fighters and their tools are recognized as an important element of the community \citep{halfakerRiseDeclineOpen2013, geigerWorkSustainingOrder2010}. 
Vandalism-related research has tended to focus on the detection and removal of vandalism \cite{tramullasResearchWikipediaVandalism2016}, with relatively little attention paid to understanding vandals themselves. Notable exceptions include
\citet{shachafVandalismWikipediaTrolls2010} who considered the cases of four people who Wikipedia community administrators (``sysops'') had identified as ``trolls'' who had engaged in acts of vandalism, including replacing main page photos with pornography and writing threats. In their interviews with sysops, Shachaf and Hara found that the sysops interpreted trolls as having engaged in harmful edits intentionally, repetitively, and in violation of policy, and that they tended to target the community itself for damage. The sysops further interpreted the trolls as being motivated by ``boredom, attention seeking, revenge....fun and entertainment....[and desire to do] damage to the community'' \cite[p. 357]{shachafVandalismWikipediaTrolls2010}.

\citet{sierraUniversityStudentsEducational2018} asked focus groups of students and Wikipedia editors to speculate as to why people vandalize. Students thought vandals might be making a joke, acting out of boredom, or engaging in ideological protest. Wikipedia editors added to these potential motives the possibility of experiencing a sense of pride at defeating Wikipedia's defenses against these vandalism. These interpretations of vandal behavior suggest that increasing the time required to vandalize may reduce more casual and sociable forms of vandalism, but that some kinds of vandals may find obstacles part of the appeal. \citet{cruzTrollingOnlineCommunities2018} was able to interview individuals identified as trolling in an online community, as well as witnesses to trolling, and found that effective trolling requires knowledge of a community's rules and norms. This suggests that to the extent that some vandalism qualifies as trolling, it may emanate from community members who know the norms well enough to intentionally transgress them. Such individuals may have more to lose if their behaviors lead to their accounts being blocked and may seek out ways to avoid identification.

There are many types of vandalism. Some have only a mildly negative impact, and others might even carry a positive impact alongside their harm. Acts of vandalism may be amusing, insightful, or comprise a political protest.
Some scholars have found evidence that unwanted behavior on Wikipedia serves as a signal of attention from the general public, which in turn may inspire further efforts or improvements from the community. \citep{gorbataiParadoxNoviceContributions2014} 
That said, just as janitorial staff scrub away even the most supportive and friendly bathroom scribble, even the most innocuous forms of vandalism are removed from Wikipedia 
to protect the integrity of the resource.

\subsection{Theoretical Framing}
One may consider vandalism on Wikipedia as an example of criminality and antinormative behavior in a broad sense and seek insight from social theories used to understand deviance. One important perspective is rational choice theory (RCT), which has found applications in economics, biology, psychology, and sociology. Although multiple articulations exist, the version of RCT I use states that people select actions which they believe will maximize their utility with consideration of their preferences and any constraints associated with that action~\cite{oppContendingConceptionsTheory1999}. I use a `wide' version of RCT which \citet{goldthorpeRationalActionTheory1998} recommends as most suitable for understanding social action and relations. The wide version incorporates an understanding that people make choices with incomplete information, subject to bounded rationality, with concern for benefits beyond material wealth (such as social approval and a positive emotional state), and nuanced models of personal preferences, traits, and fears \citep{mehlkopModellingRationalChoice2010, oppContendingConceptionsTheory1999}. 

To assist in the application of RCT to analysis of specific acts, I use Value Expectancy Theory (VET), which is derived from RCT. \citet{rikerIntroductionPositivePolitical1973} articulate value expectancy theory 
as expected utility $E$, of some behavioral alternative $a$ (identified specifically as alternative $i$ among some list of alternatives) calculated as the sum of the product of two considerations: the probability of some outcome $O$ (identified specifically as outcome $j$ among some list of outcomes 1 through $N$) and the utility valuation $U$ of that same outcome, $O_j$. 
Expected utility is thus $E(a_i) = \sum_{j=1}^N p_{ij}(O_j) U(O_j)$. With this definition in hand, we can express RCT in terms of a simple relationship among expected utilities. If the expected utility of an action $i$ exceeds the utility of some other action $j$, then rational choice predicts the selection of action $i$: $E(a_i) > E (a_j) \Rightarrow a_i$.

Utility maximization of this kind need not be assessed numerically for the proposed relationship to hold. A would-be vandal might select among alternatives iteratively, examining each possible pair in turn and choosing the preferred option. After this round-robin process, if one's thinking process is consistent, a preference ranking of activities would emerge \cite{gilboaRationalChoice2010}. In order to discern whether online vandals make rational choices consistent with VET, I examine vandalism and consider the factors relevant to VET in building up hypotheses: what constraints, what preferences, and what utility applies?

\subsection{Constraints}
Given the significance of Wikipedia, the freedom with which contributors improve or damage its content, and the substantial challenge of protecting it, we might ask what factors act as constraints against vandals. Although editing is free of charge, anyone may click the edit button, make some changes, and save them. Although this barrier sounds relatively low, relatively few people actually contribute to Wikipedia compared to the number who read it. For example, English Wikipedia, the largest of the 302 different language edition, with over 6 million articles, received over 11 billion page views in April 2020 from almost 1 billion unique devices, but the total number of unique editors in that month was only 424,155, of whom only 68,481 made five or more contributions \footnote{\url{https://stats.wikimedia.org}}. Hence it may be that even a very low level of effort is nonetheless meaningful. The Wikipedia sysops interviewed about their impression of trolls' motives by \citet{shachafVandalismWikipediaTrolls2010} included ``ease of execution'' as an important factor in troll behavior, suggesting that time constraints may be relevant. Creating an account costs more in time and energy than making the same contribution without an account because a person must fill out a brief form and choose a unique name and acceptable password.

Because VET predicts that users who have expended more effort are less likely to choose deviant behavior, and I hypothesize that \textbf{(H1) users who have created accounts will vandalize less frequently}.

A second potential constraint against vandalism is fear of detection. Given that internet activities are routinely monitored by employers, school officials, parents, internet service providers, and governments, vandals may be discouraged by a concern that they will be discovered. 
One component of being detected is being identified. 
\citet{friedmanSocialCostCheap2001a} describes the decision between using an anonymous pseudonym and a persistent identifier as ``a strategic variable'' in the interaction between participants in a cooperative online platform because it allows control over the spread of a user's reputation (p. 174).   

Wikipedia offers two types of identifiability for contributors. First, users can contribute without making an account in which case their IP address is publicly associated with their actions. Second, users can create and then log in to an account to which all subsequent contributions will be attributed.\footnote{\url{https://en.wikipedia.org/wiki/Special:CreateAccount}} 
Despite its arcane appearance as a sequence of numbers and/or letters, a user's IP address is identifiable information, and is regarded as such under privacy regulations such as the GDPR. Although individuals may not be aware of the fact, IP address can be used to identify a user's geographic location. With assistance from a service provider, the IP address can  identify the individual home or even computer. Wikipedia's interface encourages contributors to create an account, telling them that doing so will keep their IP address private from other contributors and the general public. 

In addition to choosing between these two levels of identifiability, users can take other steps to hide their identity, including naming themselves according to some pseudonym unlinked to their identity, avoiding sharing personal information online, using multiple, shared, or public IP addresses, or using an anonymity service to mask their IP address. Detecting whether users mitigating against detection is difficult.  
We can address this limitation with a unique dataset from \citet{tranTorUsersContributing2020}, who identified contributions made using the Tor anonymity service.

The Tor anonymity network allows users to access the Internet while maintaining IP privacy by relying on volunteers to transport network traffic using a method called ``onion routing.'' Typical network routing creates a traceable path of intermediate hops over which each network packet travels such that each hop knows the identity of every other node in the network. Onion routing only gives each layer---each step in the network path---the identity of the next ``hop.'' Hence, none of the intermediate steps know the identity of the point of origin, and none of them know the destination; they only know the next step~\citep{huangOnionRouterUnderstanding2016}.

Although Tor is sometimes characterized as a tool for criminals and is blocked or limited by some online services \citep{mcdonaldPrivacyAnonymityPerceived2019}, privacy-protecting tools are a crucial line of defense for people living under oppressive regimes (e.g. Iran, \cite{nazeriCitationFilteredIran2013} or Turkey \cite{akgulInternetCensorshipTurkey2015}), or whose identities or interests subject them to harassment and attacks, or who simply seek relief from surveillance and the monetization of their online behavior \cite{fortePrivacyAnonymityPerceived2017a, kangWhyPeopleSeek2013, turnerPrivacyWebExamination2003}. Although this prior work often tends to explore broadly normative behavior that is nonetheless potentially costly to the speaker, privacy techniques can certainly be used to protect high-risk anti-normative behaviors as well.
Because VET suggests that identifiability acts as a constraint on deviant behavior, I hypothesize that \textbf{(H2) populations which differ in their identifiability produce different types of vandalism, such that the least identifiable individuals are more likely to produce vandalism that has high-risk repercussions.}

\begin{figure*}
    \centering
    \includegraphics[width=6 in]{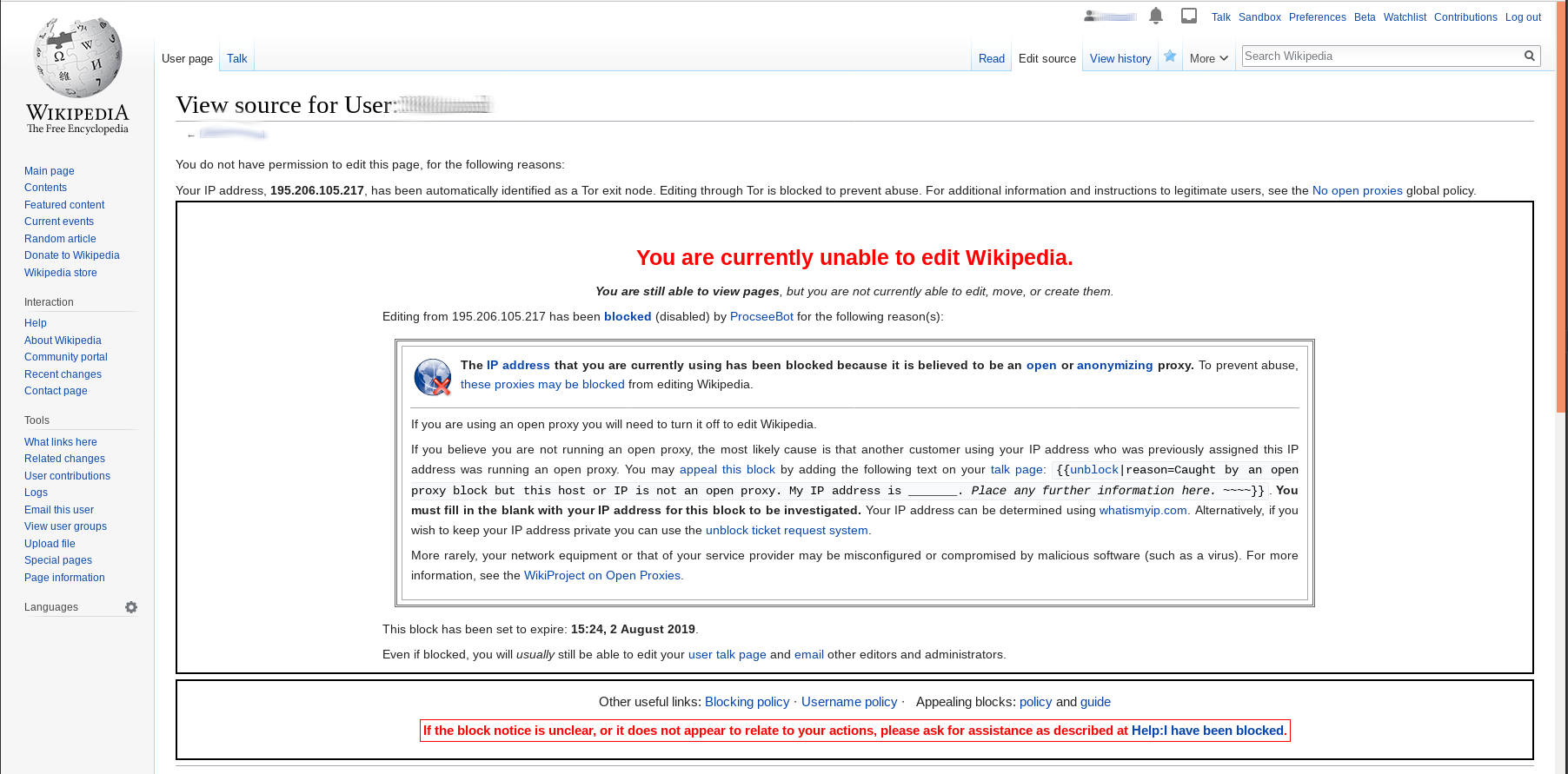}
    \caption{Regardless of whether they are logged in or not, a Tor user will see a message like the one above if they attempt to edit Wikipedia. In this case, the author is seeking to edit her own personal user page while logged in. The message states that editing one's own User Talk page and e-mailing administrators is still allowed.}
    \label{fig:torblock}
\end{figure*}

\subsection{Preferences}
The second component of a VET model is the role of preferences. In the case of vandalism, preferences can be observed in the types of vandalism a vandal chooses.
For example, if a vandal were simply desiring to do damage, any kind of vandalism might do. However, if there is some communicative intent or a specific target to be harmed, we might expect to see themes in the impact or victim/s of vandalism.  

This study compares vandals to understand whether, pursuant to \textbf{H2}, they differ in their willingness to undertake higher risk vandalism based on their expressed privacy preferences. In order to do so, I examine privacy-seekers using Tor.
Although Wikipedia forbids contributing via Tor, this policy was not always consistently or correctly applied, with the bulk of edits evading the block occurring between 2008 and 2013. During this time, about 11,000 edits were either lucky enough to fall through the cracks or able to evade the block \cite{tranTorUsersContributing2020}. 
Some of the individuals using Tor to edit Wikipedia may be privacy-seekers who simply got lucky and have no idea that Tor is blocked. Others from the sample are clearly aware of the contentious relationship between Tor and Wikipedia, and taunt Wikipedia administrators about their inability to block them \cite{championForensicQualitativeAnalysis2019}. 

As described, the groups under study differ by how they are treated by community policies. Newcomers are targeted for social interventions to welcome, train, and retain them. Wikipedia invites IP-based editors to create accounts as well as welcoming them. However, Tor-based editors generally experience rejection, despite the fact that their contributions (when they can make them) are of similar quality to newcomers and IP-based editors~\cite{tranTorUsersContributing2020}; Tor-based editors can only contribute through a combination of luck and determination. Thus it seems reasonable to examine an additional relationship between specific group traits and vandalism type. VET predicts that preferences will drive target selections, and this suggests a final hypothesis: \textbf{(H3) Members of excluded groups are more likely to strike against the community targeting them.}  

\subsection{Conceptual model}
\begin{figure*}
    \centering
    \includegraphics[width=6in]{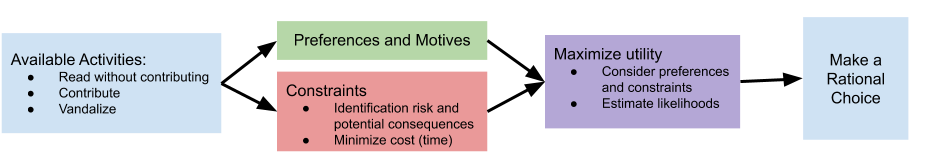}
    \caption{Conceptual model of vandalism. In the `maximize utility' stage, Value Expectancy Theory argues that individuals weigh out their possible actions against their possible outcomes, their value, and their likelihood, then choose the action they find the most optimal.}
    \label{fig:concept}
\end{figure*}
Figure \ref{fig:concept} uses VET to elaborate a conceptual model of vandalism. Read from left to right, the figure suggests that the decision to vandalize Wikipedia (as opposed to read it, contribute to it, or do nothing) begins with personal preferences and motives as well as consideration and experience of constraints. The model anticipates an individual weighing the combinations of preferences and constraints with their relative likelihoods and choosing the action that they find most suits their preferences. 

\section{Data}
\label{sec:data}

The data for this study was assembled from a random sample of article revisions made by members of four different groups of editors that differ in their identifiability and in their relative time investment in editing Wikipedia: (1) users of the Tor privacy service whose identity is fully masked that I call ``Tor-based editors'' ($n=500$); (2) users who contribute without making an account but whose IP addresses are disclosed in place of a username making it highly likely that their location can be identified \citep{liebermanYouAreWhere2009} (although they may not know this) that I call ``IP-based editors'' ($n=200$); (3) users who have made an account and who are making their first contribution that I call ``New Contributors'' ($n=200$);  and (4) users with accounts for whom a given edit is not their first that I call ``Experienced Contributors'' ($n=200$). The available population of Tor-based edits is relatively small (approximately 7,700), and the Tor-based sample is a random sample of this population. For the other three groups, I drew a random sample to time-match the Tor population on a monthly basis (e.g., if our Tor population made 30 edits in January 2009, I would randomly draw 30 edits from registered editors in January 2009, and so on.) I then drew a random sample from these time-matched samples. 

These user categories differ both in their privileges in Wikipedia \citep{tranTorUsersContributing2020} and in their privacy concerns \citep{fortePrivacyAnonymityPerceived2017a}. 
 
All IP editors have fewer privileges on the site and are automatically scrutinized at higher levels. Contributors with accounts receive progressively less scrutiny and increase in their site privileges as time passes. The Wikipedia community also elevates some members for additional administrative powers. These individuals have additional abilities to enforce norms and sanction violators, including banning them from contributing. Established contributors are also typically scrutinized less by algorithms\citep{halfakerORESFacilitatingRemediation2016}. The four groups are summarized in Table \ref{tab:constraints}.

\begin{table*}
\begin{center}
\caption{\label{tab:constraints} Constraints and preferences: vandalism has differing costs and risks for different kinds of uses. } 
\begin{tabular}{l l l l}
Editor Population & Effort Required & Privacy Risk & Privacy Awareness \\
\hline
Tor-based editors & low to high: may require several & little to none & likely to be \\  
& minutes to initiate or fail due to blocks & & privacy-aware \\
IP-based editors & none: no sign-up required & location made public & may be unaware \\
New Account & low to moderate:  registration process & location admin-visible &  may be unaware \\
Experienced Account & moderate to high: account access & location admin-visible & may be unaware \\
& may be lost or reputation damaged & & \\
\end{tabular}
\end{center}
\end{table*}

\section{Methods} 
\label{sec:methods}

This study followed a mixed-methods approach, starting with a two-stage qualitative content analysis followed by statistical analysis. 
I made use of digital trace data made public by the Tor Project and by Wikipedia, benefiting from a sample prepared in the course of 
\citeauthor{tranTorUsersContributing2020}'s \cite{tranTorUsersContributing2020}
comparison of Tor-based contributors to the same contributor groups I examine in this paper. Because the research involved only public data and did not involve interaction or intervention, the research was determined to not be human subjects research by the IRB at the University of Washington. Despite this, I have omitted names of most editors and articles and, in some cases, paraphrased quotes in order to make reidentification more difficult.

Identifying vandalism from the sample described above made use of my experience as both an editor of Wikipedia and a researcher who has observed the community for several years.
 
I defined vandalism using a three-part definition, where all three criteria must be met. First, an edit must be something which does not comply with community norms or is not encyclopedic, and hence should be removed. In the language of Wikipedia, these are ``damaging'' edits. Second, the edit must not seem to be intended to contribute to the article in a direct and constructive way. An edit can be damaging but made in good faith if it violates some rule about formatting but otherwise appears well-intentioned. Third, an edit should not be part of an ``edit war.''

An edit war involves multiple parties changing the content of a particular article back and forth. Edit wars may be due to competing visions of what is correct (e.g., moving an article back and forth between English spelling and British spelling), or due to competing beliefs about what is true (e.g., a dispute about how best to summarize the plot of a novel or the reasons for a historical event), or due to competing beliefs about the nature of knowledge (e.g., describing herbal medicine as pseudoscientific versus a reasonable and safe alternative). 
According to Wikipedia policy, disputes over content should be resolved through discussion and, if needed, arbitration---not by continually doing and undoing the work of others. Edit wars, while damaging to the article and contrary to community norms about engaging in good faith, are ultimately content-level disputes. By contrast, vandalism disrupts content and seeks to amuse, offend, disrupt, make comments to, or deliberately misinform the reader. In summary, vandalism in Wikipedia is damaging, bad-faith, and communicative or meta-communicative.  
The proposed analysis required that I identify and characterize acts of vandalism, a process which I will now describe.
The categorization schema is used is drawn from research on graffiti and vandalism in physical locations as well as specifically within Wikipedia. 
After categorizing each act of vandalism by its type, I used statistical tests to assess if observed differences were statistically significant.

\subsection{Constructing an Ontology of Vandalism}
In order to classify vandalism into types, I extend two ontologies---one from \citet{chinDetectingWikipediaVandalism2010} which describes vandalism in the context of Wikipedia, and one from \citet{whiteGraffitiCrimePrevention2001} which elaborates types of graffiti in the physical world.
My proposed synthesis is summarized with working definitions in Table \ref{fig:prelim_ontology}.

One extension in this proposed ontology is the addition of ``attack graffiti.'' While attacks on groups is combined with political graffiti by White, my contention is that while racial slurs obviously have a politics, so might many statements of opinion or fact. However, racial slurs are qualitatively different from, say, calling on the ruling party to resign, and collapsing the two together does not improve clarity. Similarly, White combines attacks on individuals with social graffiti and toilet humor. I distinguish between ``social graffiti'' including goofy nonsense and lighthearted toilet humor and ``attack graffiti'' including intimidation, insults, and threats. These distinctions are not always immediately clear and require some interpretation, just as some nuance exists between teasing and bullying.

My proposed ontology does not distinguish between what White calls ``political graffiti'' and ``protest graffiti.'' White defines protest graffiti as a subset of political graffiti targeting the content of ``mainstream commercial visual objects.'' In the context of an online encyclopedia, versus a physical environment full of advertisements and signs, a distinction singling out commercial targets is less salient. Additionally, White adopts the emic vocabulary of street graffiti to describe tagger graffiti versus gang graffiti. White describes tagging as a message of ``I'm here'' while gang graffiti asserts power and control. My proposed ontology follows this underlying distinction in the categorization scheme, but renames ``gang graffiti'' to ``pro-group graffiti'' to avoid confusion. I also omit White's category of ``Graffiti Art''. 

Applying the ontology proposed in this study is necessarily interpretive, derived from the coder's personal perspective, background, and knowledge, including social norms about gender and what qualifies as a joke. The vandalistic acts which fell within the ``attack graffiti'' category in this analysis are varied in their severity, including gross insults, repeated name substitutions, racial slurs, and rape threats against a specific Wikipedia administrator. Indeed, all categories contain variation in their severity.

\begin{table*}
\begin{center}
\caption{\label{fig:prelim_ontology}
Proposed Ontology of Vandalism}
\begin{tabular}{l l l}
Category & Description & Source \\ 
\hline
Blanking & Large content deletion & \citet{chinDetectingWikipediaVandalism2010} \\
Large-scale editing & ``Massive'' insert or change & \citet{chinDetectingWikipediaVandalism2010} \\
Misinformation & Replace content with false information & \citet{chinDetectingWikipediaVandalism2010} \\
Image attack & Insert or replace an image with an irrelevant one. & \citet{chinDetectingWikipediaVandalism2010} \\
Link spam & Adding irrelevant links & \citet{chinDetectingWikipediaVandalism2010} \\
Irregular Formatting & Insert or remove tags incorrectly & \citet{chinDetectingWikipediaVandalism2010} \\
Political Graffiti & public policy, law, power structures, social norms & combines two categories from \\
& & \citet{whiteGraffitiCrimePrevention2001}: political graffiti \\
& & and protest graffiti \\
Attack graffiti & attack an individual or group & \textbf{new}\\
Pro-group graffiti & asserting group power &  \citet{whiteGraffitiCrimePrevention2001} \\
Tagging & assertion of personal presence, bragging & \citet{whiteGraffitiCrimePrevention2001} \\
Community-related Graffiti & opposition to community, norms, or policies & \textbf{new}\\ 
Social Graffiti & Gossip, jokes, anat omical comments & called ``Toilet and Other Public''\\
&&in \citet{whiteGraffitiCrimePrevention2001} \\
\end{tabular}
\end{center}
\end{table*}

 \begin{figure*}
   \centering
    \includegraphics[width=6 in]{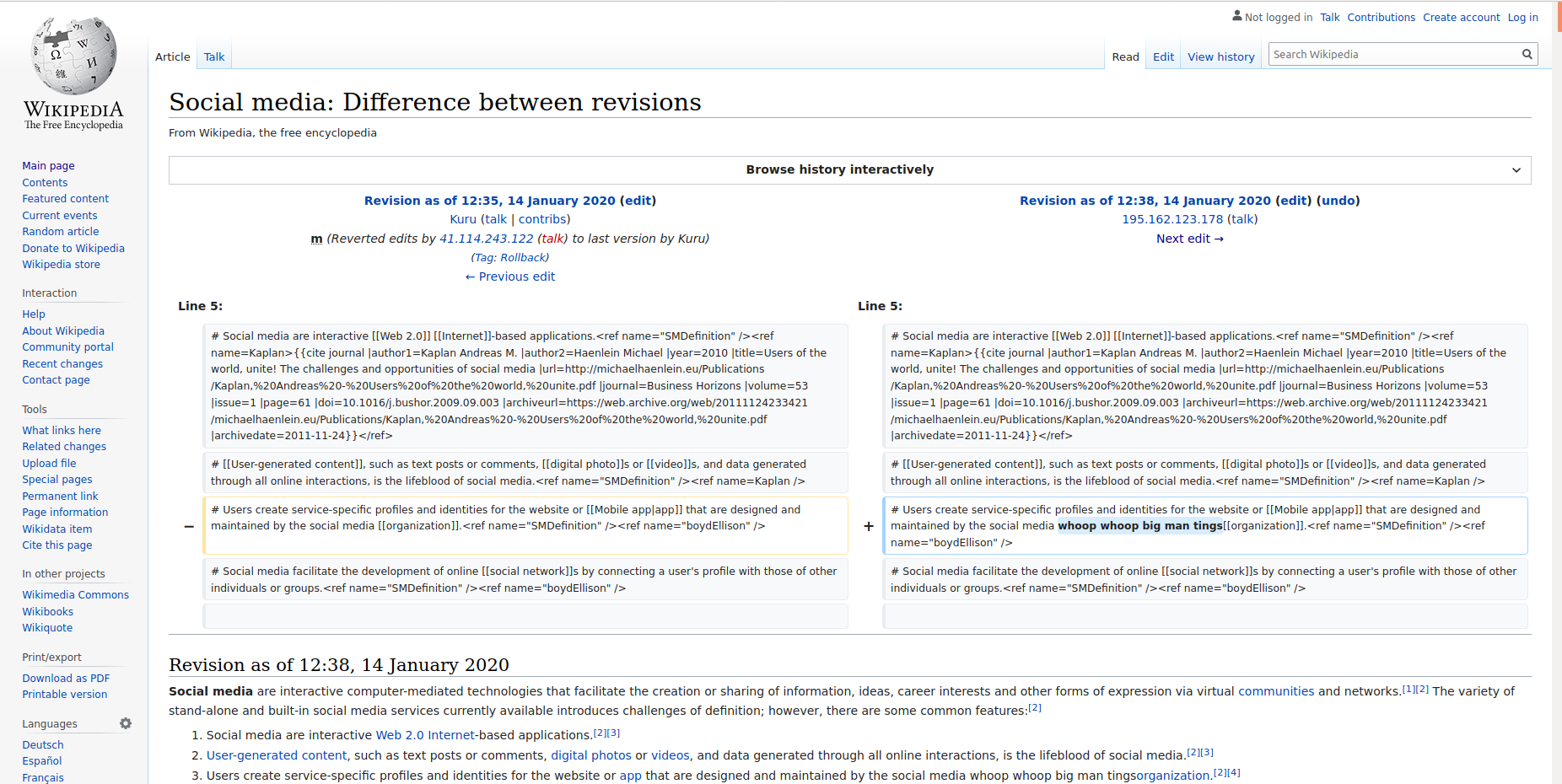}
    \caption{This screenshot is taken from Wikipedia's ``diff'' view. It highlights the differences between two versions of the article on Social Media. The left side shows the ``before'' view; ``after'' is on the right. The highlighted blue area with a + sign indicates what was added by the version. In this case, an IP-based editor adds a type of vandalism this paper would characterize as social (``whoop whoop big man tings''). Arrows above the text allow navigation through the revision history of the article.}
    \label{fig:diff}
 \end{figure*}

\begin{figure*}
    \centering
    \includegraphics[width=6 in]{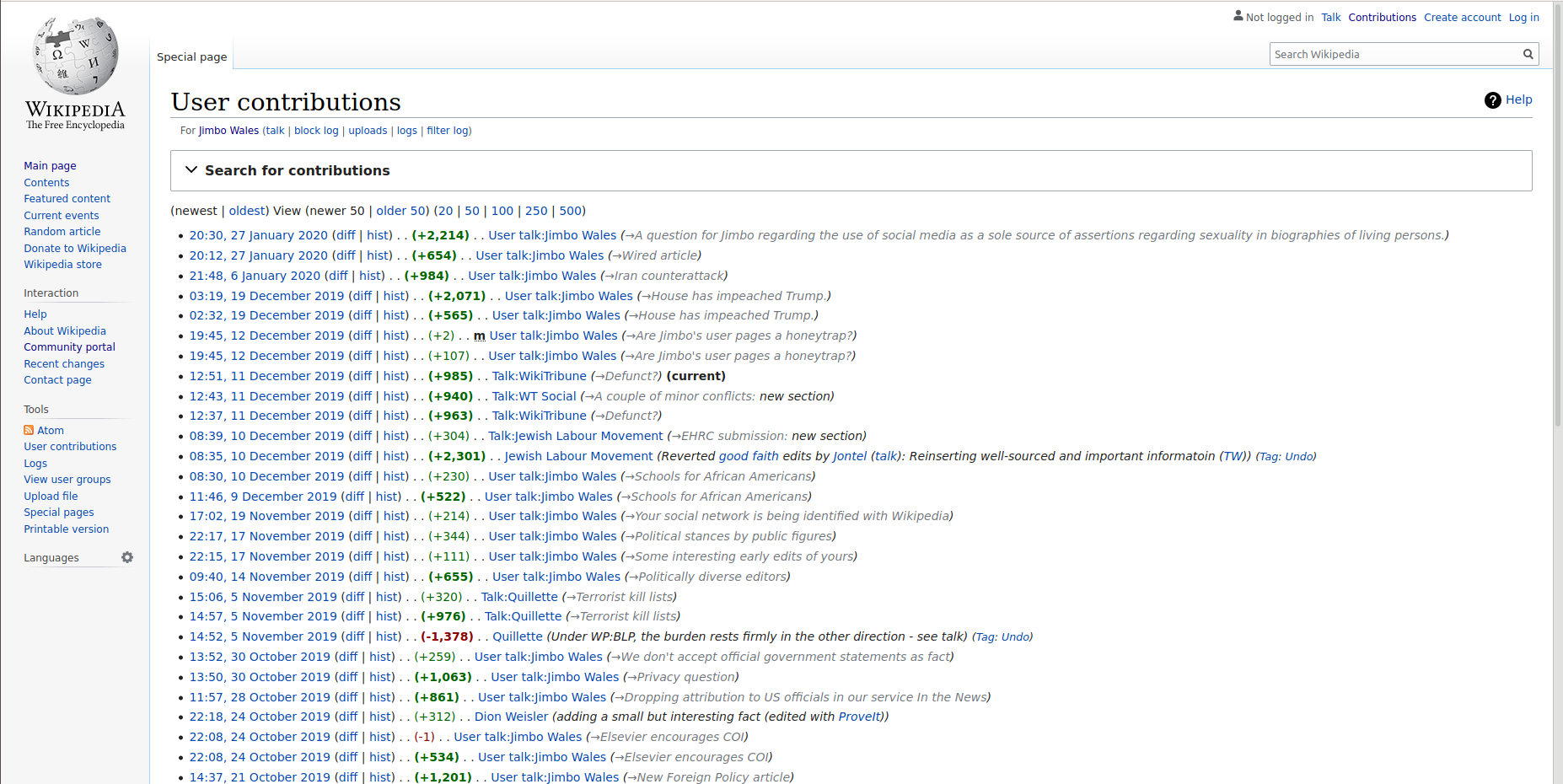}
    \caption{This screenshot is taken from a Wikipedia user's contribution page, in this case the contributions from the co-founder of Wikipedia, Jimmy (Jimbo) Wales. This information is available for all contributors.    }
    \label{fig:contrib}
\end{figure*}

Wikipedia includes a range of features for reviewing the history of articles as well as the contributions of individual users. Figure \ref{fig:diff} shows the interface that anyone can use to walk through the revision history of a given article and replay the timestamped sequence of the actions that built the page. As seen in Figure \ref{fig:contrib}, anyone can search for and view the list of contributions made by a given user. 
These features help readers, contributors, and researchers understand users' history in Wikipedia, trace interactions among contributors, and record the evolution of each page through time. 

\section{Findings}
\label{sec:results}
\subsection{Examples of Vandalism From Wikipedia}

Despite the small sample of 141 examples of vandalism found by the manual content analysis on 1100 revisions described in §\ref{sec:methods}, I identified examples of all but one of the categories in my proposed ontology of vandalism.

\textbf{Blanking} vandalism removes sections of text and entire articles. Sometimes the text is replaced, other times it is simply removed. For example, on June 19, 2010, an IP-based editor deleted a section of an article about a historical figure. The section was written in an encyclopedic tone and had several references, and the IP editor did not leave any explanation. In another example, a newly-created account using the name of a rap group deleted all of the text on the page of another rap group which appeared to be from the same US city. The lengthy text of the victimized group's page was replaced with a brief description and a link to a myspace.com page with the same name as the vandal. 

\textbf{Large-scale editing} involves adding a substantial amount of the text to the page. For example, a Tor-based editor inserted multiple paragraphs of repeated text stating over and over that ``[Celebrity stalker] died.'' in the text of several television and celebrity articles on December 24, 2008. In other cases, the text added by a Tor-based editor is varied and seemed to be making an argument, but not one relevant to the page. One example involved a Tor user critiquing an administrative decision in Wikipedia, claiming to be an individual who had been identified and banned for misusing dozens of accounts (a practice called ``sockpuppeting''), and calling other editors ``evil'' on multiple occasions from February to April 2008.

\textbf{Misinformation} vandalism refers to attempts to deliberately introduce false information or to assert information based on rumors. One example of misinformation vandalism occurred on February 26, 2013 when a Tor-based editor changed two letters in a word in a long quotation, changing the meaning of the quote. I looked at multiple primary sources and all of them contradicted this change. This misinformation persisted until November 5, 2013. Another incident of misinformation vandalism occurred on February 1, 2013, when an IP-based editor altered the biography of a broadcaster stating that he had been shot to death in his home. News reports contemporary to the edit described his death as occurring in a hospital after an illness. Wikipedia includes an option for editors to summarize their change as a note for other editors to see. In this case the vandal described their change as fixing a ``minor spelling error.'' However, their edit did not change any spellings.

\textbf{Image attack} vandalism involves substituting an irrelevant or harassing image for the legitimate image on a page. This type of vandalism is difficult to assess retrospectively because generally a removed image is no longer available online. 
In some cases, the nature of the attack is clearer: for example, on September 23, 2009, a new editor replaced a female politician's photo with an image with the word ``horse'' in the name, which was reversed 33 minutes later. 

\textbf{Link spam} is the insertion of links to external sites in violation of the Wikipedia external links policy. These links may be commercial in nature or connected to malware sites. For example, on February 11, 2012, a Tor-based editor inserted a link to a spam website into an article about a card game. The links were removed 7 days later as part of a group of edits removing similar links.

\textbf{Irregular formatting} refers to vandalism which specifically targets the wiki-specific syntax of a page. This includes special code-like symbols such as square brackets [], curly braces \{\}, and so on which control how the page appears. Altering a single symbol can cause the entire page to display incorrectly or to be unreadable.  I observed this type of vandalism in two cases: for example, on December 17, 2012, an IP-based editor removed ``\}\}'' from a biography of a prominent female scientist, causing the top portion of the page to appear jumbled and difficult to read. 

\textbf{Political graffiti} is vandalism that invokes or protests public policy, law, power structures, and social norms. I observed only one example of this type of vandalism in which an IP-based editor on December 13, 2011 added to an article which described a particular substance as having been historically believed to diminish homosexual desire stating that ``these...insane theories still persist to the current day'' and that ``People need to learn to mind their own [expletive] business.'' This mostly empty category suggests further refinement to the ontology is needed. 

\textbf{Attack graffiti} insults, seeks to intimidate, or seeks to humiliate individuals or groups. One example of attack graffiti occurred on September 20, 2008 when a newly-created account edited a Wikipedia administrator's personal messages page (called their ``User Talk'' page within Wikipedia) to include a description of the administrator being raped by an individual who had recently been banned. I found that threats of this kind were made repeatedly from both IP-based user identities and newly created accounts. The administrator used a female-presenting name and the vandal in question used a male-presenting name. Another example of attack graffiti occurred on December 31, 2007, and altered a page about medieval history with text which altered the word Wikipedia to contain a reference to pedophilia and then to state that they should ``...should probably start blocking Tor.'' This edit was also coded as community-related graffiti.

\textbf{Pro-group graffiti}, the redefinition of ``gang'' graffiti for the context of Wikipedia is graffiti which asserts some group identity or ownership. The sample as coded did not contain any examples of this type of graffiti, suggesting that the ontology may need to be refined in this dimension as well.

\textbf{Tagging} in the context of Wikipedia is defined as actions which assert personal presence or brag in some way about an individual, presumably oneself. Of course, it cannot be determined if tagging is perpetrated by the person whose name is used or if it is done by a rival seeking to draw negative attention to an apparent braggart. One example of tagging graffiti in Wikipedia occurred on an article about a beer company, in which on February 5, 2008 a newly registered account added the phrase ``[name] is a pimp'' in capital letters. Another example of tagging graffiti occurred on December 17, 2007, when an IP-based editor altered a page about an online first-person-shooter video game to include five sentences describing a particular player by name as being a dominant player in the game, ending with the phrase that he ``pwns noobs out da club.''

\textbf{Community graffiti} is vandalism that references the specific community where it appears. In my case, this type of vandalism referred directly to Wikipedia or its policies, or made use of sophisticated understanding of Wikipedia features to target Wikipedia administrators with vandalism.
One example occurred on April 30, 2008 when a Tor-based editor repeatedly updated the user page associated with their own Tor IP in such a way as to
trigger a malfunction in an automated script (``bot'') used by the Wikipedia community to monitor and block Tor-based users. The malfunction persisted until November 10, 2008 when a different bot cleaned up the page.

\textbf{Social graffiti} refers to joking comments, anatomical references, and exchanges of greetings. One example of social graffiti occurred on December 9, 2007, when an IP-based user altered a page about a technology to include the words ``stupid people are so stupid'' and some random characters. Another example occurred on December 3, 2011, when a newly-created account updated an article about a particular species of plant to claim that it grows in a fictional videogame universe.

\subsection{Vandalism Rate}

The relative prevalence of vandalism in each group is summarized in Table \ref{tab:vanRates}. I observed vandalism in 13\% of the Tor-based edits, 24\% of the IP-based edits, 1\% of registered user edits, and 15\% of the new editor edits. In order to understand whether the variation in rates by group is statistically significant, I used a $\chi^2$ test of a frequency table containing the counts of vandalism and non-vandalism for each type. The $\chi^2$ reported $p < 0.001$, indicating that the difference among user types is statistically significant. 
 
As a robustness check, I verified that the results of the $\chi^2$ test of the statistical significance of the differences between vandalism rates still held without considering registered editors (pursuant to \textbf{H1}); this result was confirmed with a p-value estimated at $p < .002$.
This provides evidence in support of \textbf{H1} that populations which have expended more effort will vandalize less frequently.

\begin{table}
\begin{center}
\caption{\label{tab:vanRates} The rate of vandalism in the sample.}
\begin{tabular}{l l l l}
Editor Population & Vandalism Prevalence \\
\hline
Registered Editors & 1\% \\
First-time Editors & 15\% \\
IP-based Editors & 24\% \\
Tor-based Editors & 13\% \\
\end{tabular}
\end{center}
\end{table}

The highest-effort group (registered editors) engaged in almost no vandalism. Tor-based editors arguably must work harder than IP-based editors, and their rate of vandalism was lower. This observed variation in behavior by user types is somewhat surprising given prior work which found that levels of non-damaging editing behavior are roughly equivalent among the first-time, IP-based, and Tor-based editor populations \cite{tranTorUsersContributing2020}.

The full sample of revisions examined through content analysis ($n=1100$) was used for assessing \textbf{H1}. However, because the rate of observed vandalism was so low in registered editors (2 instances in the sample of 200 edits), further analysis was impractical. The registered editor group is omitted from consideration in \textbf{H2} and \textbf{H3}, and the only the identified acts of vandalism in the other three groups ($n=141$) are used for these latter two hypothesis tests.

\subsection{Vandalism Types}
To test \textbf{H2}, that types of vandalism will vary such that groups with lower identifiability will be more likely to engage in vandalism, I determined the relative prevalence of vandalism types. 

Table \ref{tab:freq} reports both the average incidence rate of types of vandalism and whether or not the differences in the reported means are statistically significant as assessed at the $\alpha=.05$ level via a one-way ANOVA. 
I find partial support for \textbf{H2}. Groups that vary in their identifiability also vary in their vandalism type, with low identifiability associated with higher risk types of vandalism. Tor-based users are substantially more likely than other groups to engage in large-scale vandalism and least likely to engage in the lowest risk type of vandalism, that which communicates friendly and sociable intent. Although other potential high-risk types of vandalism, such as blanking, misinformation, link spamming, or attacking people, was higher among Tor-based editors, this difference was not statistically significant.

This analysis also supports \textbf{H3}. Excluded groups are more likely to vandalize in ways that strike against the community: Tor-based editors target the community itself at a much higher rate as seen in Table \ref{tab:freq}.

\begin{table*}
\begin{center}
\caption{\label{tab:freq}Relative prevalence of observed vandalism types. Percentages do not sum to 100\% because multiple codes were applied to individual instances of vandalism (n=141). \textbf{Bold text} indicates that the difference in means among the groups is statistically significant at the .05 level as determined via a one-way ANOVA. Registered editors were eliminated as a comparison group due to their low overall incidence rate of vandalism in the sample. No instances of ``pro group'' graffiti were identified in the sample, hence the category does not appear here.}
\begin{tabular}{l l l l l l l l l l l l l}
\hline
\textbf{Group} & \emph{Blank} & \textbf{Large} & \emph{Misinfo.} & \emph{Image} & \emph{Link} & \emph{Format} & \emph{Political} & \emph{Attack} & \textbf{Wikipedia} & \textbf{Social} & \emph{Tag}\\
\hline
\textbf{IP-based} & 13\% & \textbf{4\%} & 9\% & 0\% & 0\% & 2\% & 2\% & 26\% & \textbf{0\%} & \textbf{55\%} & 9\% \\
\textbf{Tor-based} & 19\% & \textbf{14\%} & 13\% & 2\% & 11\% & 2\% & 0\% & 36\% & \textbf{17\%} & \textbf{19\%} & 9\% \\
\textbf{First-time} & 13\% & \textbf{0\%} & 7\% & 3\% & 7\% & 0\% & 0\% & 17\% & \textbf{3\%} & \textbf{57\%} & 17\% \\
\hline
\end{tabular}
\end{center}
\end{table*}

\section{Limitations \& Future Work}
\label{sec:limitations}
This study is limited in important ways. First, I rely on a qualitative coding process conducted by a single investigator. Of course, assessment of vandalism in Wikipedia is often a series of judgment calls from vandal-fighters working at high speed on their own. I am familiar with both the empirical context and several of the results are consistent with independent findings in other work. 

Future work should not only use a larger number of observations and a multi-investigator coding process but also should consider additional potential constraints, such as the effort required to enact a specific act of vandalism (e.g., by counting keystrokes). 

Additionally, while I have assessed each act of vandalism independently and from a random sample, some vandalistic acts are committed by serial vandals. These serial vandals may learn from the community response and adjust their approach in order to make their actions more effective. For example, 
\citeauthor{matsuedaDeterringDelinquentsRational2006}'s \cite{matsuedaDeterringDelinquentsRational2006} study of at-risk and delinquent juveniles found evidence for the influence of learning. Future research on social deviance in Wikipedia looking for evidence of learning might yield interesting results. For example, a vandal seeking to spread misinformation about a celebrity death might observe that their efforts were immediately reversed and fine-tune their language to avoid words which trigger automated vandalism detection systems, or make use of conventions which build the confidence of other editors, such as citing phony sources. 

This work is also limited because the study does not include direct discussion with vandals. Past studies of crime have benefited tremendously from interviewing, for example at-risk adolescents as in \citet{matsuedaDeterringDelinquentsRational2006}, or people observed as trolling in an online community as in \citet{cruzTrollingOnlineCommunities2018}. Although such conversations might be enlightening, the generally anonymous Wikipedia vandal population might be tremendously difficult to contact since no contact or communication mechanism is required in order to contribute to Wikipedia. 
Finally, this study makes only a preliminary use of RCT and VET and a future investigation could both ground the phenomena more deeply in these perspectives as well as develop expanded empirical models.

\section{Discussion}
\label{sec:discussion}

Taken together, the results for my three hypotheses add texture to the more general explanation often offered for antinormative behavior online in terms of online disinhibition: the notion that being more distant and less identifiable may reduce one's reluctance to give in to one's worst impulses \cite{sulerOnlineDisinhibitionEffect2004, sulerBadBoysCyberspace1998, joinsonCausesImplicationsDisinhibited1998}. Instead, we see that the least identifiable group (Tor-based editors) does not have the highest rate of vandalism.  The supportive evidence found for \textbf{H1}, \textbf{H2}, and \textbf{H3} reinforce the utility of applying Rational Choice and Value Expectancy theories and suggest that identifiability and time may serve as operative constraints on vandal behavior.

One way to use these results to counter vandalism is to consider how potential interventions might change the kinds of vandalism a community receives. For example, adding a CAPTCHA \cite{vonahnCAPTCHAUsingHard2003} to slow down contributions increases costs for would-be contributors. However, given that the two groups for whom contributing is most costly (new editors and Tor-based users) seem to have a different pattern of vandalism types than other groups, increased time requirements alone may not change the rate or quality of the vandalism. \citet{friedmanSocialCostCheap2001a} proposed an alternative approach to the question of deterring antinormative behavior without requiring identifiability. They suggested anonymous certificates, using cryptographic techniques both to protect identity and to guarantee a single individual will only have one anonymous certificate. Another intervention might be to increase the sophistication of tools that detect and filter contributions before they become visible: for example, disallowing some groups from making large-scale edits or blanking pages might be desirable in some communities. Tools seeking to detect vandalism may find that the patterns that can be identified within each category allow for more accurate targeting algorithms.
Although the differences in the prevalence of several other categories in this sample were not statistically significant, a larger-scale analysis or additional analytic variables to represent preferences and constraints may reveal additional trends.

For privacy advocates seeking policy changes, anti-community vandalism 
poses a difficult challenge. 
Although frustration from banned users is not surprising, targeting the community is disruptive and the experiences of harassment victims should not be discounted. 
Given that many of Wikipedia's defense mechanisms against damage make use of IP-based identification to block abusers as precisely as possible, accommodating those who seek IP privacy poses a substantial challenge. However, viewing vandals as rational and having diverse motives opens potential avenues for dialogue.

\section{Conclusion}
\label{sec:conclusion}

This study characterized vandalism in Wikipedia using Rational Choice Theory and the specific considerations of Value Expectancy Theory. Based on consideration of preferences and constraints, this study
assessed the types and prevalence of vandalism and found that both vary among groups of vandals in ways that are predicted by hypotheses drawn from RCT and VET. 

These findings contribute to the emerging understanding of online norm violations that are low-risk to violators but harmful to victims, communities, and public goods. Identifiability and effort, together with exclusionary policies that affect some vandals, offer some explanation for both the rates of vandalism and types of vandalism perpetrated. Interventions that target these factors independently may have unintended consequences and could deter newcomers and valuable casual contributions \citep{halfakerRiseDeclineOpen2013, halfakerDonBiteNewbies2011}.

This study contributes to the study of online vandalism by applying theories of Rational Choice and Value Expectancy in order to develop a framework for understanding and modeling vandal behavior. Doing so required an extension and synthesis of two vandalism ontologies and bringing together perspectives rooted in online and offline vandalism. This work suggests that treating vandalism as the product of an antinormative but rational decision-making process supports further consideration of different strategies for countering and preventing it.

\subsection{Acknowledgements} I am tremendously grateful for the guidance of Karl-Dieter Opp; all errors are mine, but his encouragement and mentorship were invaluable. Three anonymous reviewers provided helpful suggestions and the resulting manuscript is much improved by their feedback. This work would not have been possible without the support of Chau Tran, who kindly shared his dataset of Tor users with me and assisted in sample preparation. This work was born out of early discussions with Nora McDonald, Stephanie Bankes, and Joseph Zhang as we puzzled out what qualified as vandalistic behavior in Wikipedia. This work was supported by the National Science Foundation (awards CNS-1703736 and CNS-1703049).
\balance
\bibliographystyle{ACM-Reference-Format}
\bibliography{bibliography}
\end{document}